\input harvmac
%
\rightline{WIS/97/6, EFI-97-09, RI-1-97}
\Title{
\rightline{hep-th/9702014}
}
{\vbox{\centerline{Branes and N=1 Duality in String Theory}}}
\medskip
\centerline{\it Shmuel Elitzur, Amit Giveon}
\centerline{Racah Institute of Physics}
\centerline{The Hebrew University}
\centerline{Jerusalem, 91904, Israel}
\vskip .2in 
\centerline{\it David Kutasov\footnote{*}
{On leave of absence
from Department of Physics, University of Chicago,
5640 S. Ellis Ave., Chicago, IL 60637, USA.}}
\smallskip
\centerline{Department of Physics of Elementary Particles}
\centerline{Weizmann Institute of Science}
\centerline{Rehovot, 76100, Israel}
\vglue .3cm
\bigskip
 
\noindent
We propose a construction of dual pairs in four dimensional
N=1 supersymmetric Yang-Mills theory using branes in type IIA 
string theory.

\Date{2/97}

\def\journal#1&#2(#3){\unskip, \sl #1\ \bf #2 \rm(19#3) }
\def\andjournal#1&#2(#3){\sl #1~\bf #2 \rm (19#3) }

\def\ie{{\it i.e.}}
\def\eg{{\it e.g.}}

\def\frac#1#2{{#1\over#2}}

\def\inbar{\,\vrule height1.5ex width.4pt depth0pt}
\def\IC{\relax\hbox{$\inbar\kern-.3em{\rm C}$}}
\def\IR{\relax{\rm I\kern-.18em R}}
\def\IP{\relax{\rm I\kern-.18em P}}

%
%
\def\np#1#2#3{Nucl. Phys. {\bf B#1} (#2) #3}
\def\pl#1#2#3{Phys. Lett. {\bf #1B} (#2) #3}

\catcode`\@=11
\def\slash#1{\mathord{\mathpalette\c@ncel{#1}}}
\overfullrule=0pt

\def\underrel#1\over#2{\mathrel{\mathop{\kern\z@#1}\limits_{#2}}}

\catcode`\@=12


%


\def\nsp{{NS$^\prime$}}


\newsec{Introduction.}

\nref\neqone{N. Seiberg, hep-th/9408013; 
hep-th/9506077.}%
\nref\nonerev{K. Intriligator and N. Seiberg, 
hep-th/9509066.}%
Supersymmetric gauge theories invariant under 
eight or sixteen supercharges (\eg\ N=2 and N=4
supersymmetric Yang-Mills theories in four dimensions,
respectively) have been intensively studied in the
last few years. Many of the diverse strong coupling
phenomena that they exhibit were found to have a
natural explanation in string theory (some were even
discovered this way). In gauge theory, a lot is known
about systems with four supercharges (in particular,
N=1 supersymmetric gauge theories in four dimensions
\refs{\neqone, \nonerev}), 
and it is natural to ask what string 
theory can teach us about such systems.

The purpose of this note is to construct, in analogy to
\ref\hw{A. Hanany and E. Witten, hep-th/9611230.},
a configuration of branes in type II string theory,
which appears to exhibit one of the most remarkable
field theory phenomena -- Seiberg's duality in
four dimensional N=1 supersymmetric gauge theory 
\ref\nati{N. Seiberg, hep-th/9411149, 
\np{435}{1995}{129}.}.
This duality is the statement that 
two different gauge
theories (``electric'' and ``magnetic'') 
may sometimes
give rise to the same long distance physics, 
\ie\ flow to the
same infrared conformal field theory. 
Its interpretation
in string theory was recently studied in 
\ref\vvv{M. Bershadsky, A. Johansen, 
T. Pantev, V. Sadov and C. Vafa, 
hep-th/9612052; C. Vafa and 
B. Zwiebach, hep-th/9701015.}
from a different point of view.
We will also discuss certain 
extensions of Seiberg's work
\ref\ks{D. Kutasov, hep-th/9503086, 
\pl{351}{1995}{230};
D. Kutasov and A. Schwimmer, 
hep-th/9505004, \pl{354}
{1995}{315}; D. Kutasov, A. Schwimmer and N. Seiberg,
hep-th/9510222, \np{459}{1996}{455}.}.

Below, we consider a configuration of branes whose
low energy worldvolume dynamics is that of an N=1
four dimensional 
supersymmetric Yang-Mills theory with gauge group 
$U(N_c)$ and $N_f$ flavors of quarks in the fundamental
representation. We focus on two microscopic coupling
constants, the gauge coupling and
the Fayet-Iliopoulos (FI) D-term. 
We show that by varying the gauge coupling 
(using the FI D-term to avoid a strong
coupling singularity), we recover
in different limits the electric and magnetic 
descriptions of the theory, given by Seiberg. 
This shows that the two models have the same
moduli space of vacua and, therefore, the same
chiral ring. It is natural to expect that the
full infrared conformal field theories agree as
well.

\bigskip

\newsec{The brane configuration.}

The configurations we will study involve four
kinds of branes in type IIA string theory: 
a Neveu-Schwarz (NS) fivebrane,
Dirichlet (D) sixbrane, Dirichlet 
fourbrane and a differently oriented NS fivebrane, 
which we will refer to as the \nsp\ fivebrane. 
Specifically, the four kinds of branes are:

\item{(1)} NS fivebrane with worldvolume 
$(x^0, x^1, x^2, x^3, x^4, x^5)$, which lives at 
a point in the $(x^6, x^7, x^8, x^9)$ directions.
The NS fivebrane preserves supercharges of the 
form\foot{$Q_L$, $Q_R$ are the left and right moving 
supercharges of type IIA string 
theory in ten dimensions. They are (anti-) chiral: 
$\epsilon_R=-\Gamma^0\cdots\Gamma^9\epsilon_R$,
$\epsilon_L=\Gamma^0\cdots\Gamma^9\epsilon_L$.}
$\epsilon_LQ_L+\epsilon_RQ_R$, with 
\eqn\nsfive{
\eqalign{\epsilon_L=&\Gamma^0\cdots\Gamma^5\epsilon_L\cr
\epsilon_R=&\Gamma^0\cdots\Gamma^5\epsilon_R.\cr
}}

\item{(2)} D sixbrane with worldvolume
$(x^0, x^1, x^2, x^3, x^7, x^8, x^9)$, which lives at a point
in the $(x^4, x^5, x^6)$ directions.
The D sixbrane preserves supercharges satisfying
\eqn\dfive{
\epsilon_L=\Gamma^0\Gamma^1\Gamma^2\Gamma^3\Gamma^7\Gamma^8
\Gamma^9\epsilon_R.
}

\item{(3)}
D fourbrane with worldvolume $(x^0, x^1, x^2, x^3, x^6)$
which preserves supercharges satisfying 
\eqn\dthree{
\epsilon_L=\Gamma^0\Gamma^1\Gamma^2\Gamma^3\Gamma^6\epsilon_R.
}

\item{(4)} \nsp\ fivebrane with worldvolume
$(x^0, x^1, x^2, x^3, x^8, x^9)$ preserving the
supercharges
\eqn\nsprime{
\eqalign{\epsilon_L=&\Gamma^0\Gamma^1\Gamma^2\Gamma^3
\Gamma^8\Gamma^9\epsilon_L\cr
\epsilon_R=&\Gamma^0\Gamma^1\Gamma^2\Gamma^3
\Gamma^8\Gamma^9\epsilon_R.\cr
}}

It is easy to check that there are four supercharges
satisfying equations \nsfive-\nsprime, 
1/8 of the original supersymmetry of 
type IIA string theory.

Similarly\foot{T duality in the $x^3$ direction
turns our construction into a three dimensional
N=2 supersymmetric
version of \hw.} to \hw, we will study the 
dynamics on the worldvolume of fourbranes stretched 
between fivebranes. The case of interest will be a 
configuration of $N_c$ D fourbranes
stretched between an NS fivebrane and 
an \nsp\ fivebrane,
along the $x^6$ direction. Thus, the 
worldvolume of the
D fourbrane is $R^{3,1}$ times a finite interval $I$. 
The worldvolume dynamics describes at long distances
an N=1 supersymmetric theory in 3+1 dimensions. 

Of course, to have the possibility of 
stretching a D fourbrane
between the NS and \nsp\ fivebranes without 
breaking supersymmetry, 
the two NS branes must coincide in the $x^7$ 
direction. We will see 
later what happens when they do not coincide.
 
Since the configuration of a D fourbrane 
between NS and \nsp\ fivebranes preserves four
supercharges (even in the absence of D sixbranes),
it describes a 3+1 dimensional N=1 supersymmetric 
gauge theory at low energies. It is not difficult 
to deduce the field content. The fourbrane 
worldvolume fields that describe its
fluctuations in the $(x^6, \cdots, x^9)$ 
directions are fixed by the 
boundary conditions at the NS fivebrane end. 
Those that describe 
fluctuations in the $(x^4, \cdots, x^7)$ 
directions are fixed by 
boundary conditions at the \nsp\ fivebrane end. 
Thus, the only
massless mode arising from 4-4 strings 
is the $U(N_c)$ worldvolume gauge field.

To add matter, we insert (as in \hw) 
$N_f$ D sixbranes
at values of $x^6$ that are between the 
positions of the NS and \nsp\ fivebranes. 
The 4-6 strings describe $N_f$ chiral multiplets 
in the fundamental representation of $U(N_c)$. 
Note that:

\item{1)} The relative 
position of the NS and \nsp\ 
fivebranes in the $x^7$ 
direction plays the role of a 
FI D-term in the $U(N_c)$ 
gauge theory on the
fourbranes. To find a 
supersymmetric vacuum when it is
non-vanishing, one has to turn 
on Higgs expectation values
for the quarks. This can be 
achieved by a similar mechanism 
to that described in \hw, 
whereby a D fourbrane stretched 
between the NS and \nsp\ fivebranes 
touches a D sixbrane
(at which point the mass of the 
corresponding chiral multiplet
vanishes), and splits into two 
fourbranes, one stretched between
the NS and D sixbranes, and the 
other between the D sixbrane
and the \nsp\ fivebrane.

\item{2)} The distance between 
the NS and \nsp\ fivebranes
in the $x^6$ direction, $L_6$, 
determines the 3+1 dimensional
fourbrane worldvolume 
gauge coupling: $1/g_4^2\propto L_6$.

\item{3)} The distance in the $(x^4, x^5)$ directions
between the \nsp\ fivebrane and the $N_f$ D sixbranes
determines the masses of the $N_f$ chiral multiplets.

\item{4)} It is possible for a fourbrane stretched
between the NS and \nsp\ fivebranes to break on a
D sixbrane into two pieces 
with a relative splitting in the
$(x^8, x^9)$ directions. This corresponds to turning
on a Higgs expectation value for one of the quarks.

To summarize, the brane configuration we 
start with, viewed along
the $x^6$ direction, is the following: 
the NS fivebrane is the leftmost
object, and is connected to the \nsp\ 
fivebrane by $N_c$ fourbranes.
The D sixbranes are placed between 
the NS and \nsp\ branes. This configuration 
corresponds to the electric 
theory of \nati. In the next section we will  
describe the magnetic one. 
 
\bigskip

\newsec{Seiberg's duality.}

The dynamics on the fourbrane 
worldvolume describes an N=1 
supersymmetric $G_e=U(N_c)$ gauge 
theory coupled to $N_f$ fundamental 
``quarks'' in 3+1 dimensions. 
Seiberg's duality \nati\ is the statement 
that at long distances this system has 
another description, where the gauge 
group is replaced by $G_m=
U(N_f-N_c)$, and in addition to the 
$N_f$ chiral superfields
in the $\bf{N_f-N_c}$ of 
$U(N_f-N_c)$ there appears a $G_m$
singlet field $M$, in the 
$\bf{N_f^2}$ of the $U(N_f)$ 
global symmetry group (here this is the 
D sixbrane gauge group,
which gives a global symmetry on the fourbrane
worldvolume). The ``magnetic meson field'' 
$M$ couples to the
magnetic quarks via a superpotential 
which will be described later.

To find this ``magnetic'' description, 
we follow \hw\ and move the 
NS fivebrane to the other side of the  
\nsp\ fivebrane in the $x^6$ direction. 
This motion of the NS fivebrane in the 
$x^6$ direction corresponds to changing
the microscopic gauge coupling of the theory, 
and thus should not change the infrared behavior. 
There is a potential singularity corresponding 
to the point where the NS and \nsp\ fivebranes 
coincide (and the coupling in the gauge theory 
diverges), but that can be avoided (for $N_f\ge N_c$) 
by switching on and off the FI D-term as we vary 
the gauge coupling or, equivalently, going around 
the \nsp\ fivebrane in the $(x^6, x^7)$ plane.
As we shall see, after the NS fivebrane completes 
its motion, the system one finds is the magnetic 
description of the theory.

To facilitate the presentation, 
it is convenient to think of the 
motion described above as composed 
of the following four steps:

\item{(a)} The NS fivebrane moves to the 
right in the $x^6$ direction, crossing all 
$N_f$ D sixbranes.

\item{(b)} The NS fivebrane moves away from the 
\nsp\ fivebrane in the $x^7$ direction.
 
\item{(c)} The NS fivebrane moves in the $x^6$ 
direction, to the other side of the \nsp\ fivebrane.

\item{(d)} The NS fivebrane moves in the $x^7$ 
direction back to its original $x^7$ position.

We will next discuss in turn the steps (a) -- (d), 
focusing on the different phenomena that take place. 

\noindent
{\it Step (a):} Whenever the $x^6$ location 
of the NS fivebrane passes through that of a D sixbrane, 
the two branes actually meet  in space\foot{In contrast, 
the \nsp\ fivebrane and D 
sixbrane can avoid each other in the 
$(x^4, x^5)$ directions.}. This leads \hw\ to the 
creation of a new fourbrane connecting the NS fivebrane 
(now on the right of the D sixbrane) and the D sixbrane. 
Thus at the end of step (a), in addition to the $N_c$ 
fourbranes connecting the NS and \nsp\  fivebranes, there 
are $N_f$ fourbranes connecting the NS fivebrane to
the $N_f$ D sixbranes (all of which are now to the left 
of the NS fivebrane).

\noindent
{\it Step (b):} As mentioned above, 
the $N_c$ fourbranes stretched between
the NS and \nsp\ fivebranes preserve 
supersymmetry only when the $x^7$
positions of the two branes are the 
same. Since we now want to move the
NS fivebrane relative to the \nsp\ 
fivebrane in the $x^7$ direction, in order
not to break SUSY the gauge theory 
must be in a Higgs phase, which in the
brane language means the following. 
If the D sixbranes are at the same 
$(x^4, x^5)$ as the \nsp\ fivebrane 
(which means that the electric quarks
$Q$ are massless), then the $N_c$ 
fourbranes connecting the NS and \nsp\ 
fivebranes can connect to $N_c$ of the $N_f$ 
new fourbranes created in
step (a) and together leave
the NS fivebrane in the inverse of 
the process described in \hw, Figure
3a. Of course, for this to be possible, 
one must have $N_f\ge N_c$, a constraint 
related both to the fact 
\ref\ads{I. Affleck, M. Dine and N. Seiberg,
\np{241}{1984}{493}.} 
that SQCD has no supersymmetric vacuum for $N_f<N_c$,
and to similar observations in \hw\ 
about the theory with eight supercharges.
At the end of step (b) we thus find the 
following situation. There are $N_c$ fourbranes 
connecting the \nsp\ fivebrane to $N_c$
D sixbranes. The other $N_f-N_c$ D sixbranes 
are connected by fourbranes to the NS fivebrane. 

\noindent
{\it Step (c):} Since the NS and \nsp\ fivebranes 
do not meet in space (even when their $x^6$ values 
coincide), nothing special is expected to happen 
at this stage. 

\noindent
{\it Step (d):} As the NS fivebrane 
comes back to its original $x^7$ 
position, the $N_f-N_c$ fourbranes 
connecting it to $N_f-N_c$ different
D sixbranes touch the \nsp\ fivebrane. 
They then split into $N_f-N_c$
fourbranes stretched between the NS and 
\nsp\ fivebranes, and $N_f-N_c$
fourbranes stretched between the \nsp\ 
fivebrane and $N_f-N_c$ D sixbranes.

The final brane configuration is the following. 
The NS fivebrane is connected to the \nsp\ fivebrane 
by $N_f-N_c$ fourbranes. The \nsp\ 
fivebrane is further connected
by $N_f$ fourbranes to the 
$N_f$ D sixbranes.
To see that this is the magnetic
description of the original
theory, note that:

\item{1)} The gauge group on the fourbrane 
worldvolume is $G_m=U(N_f-N_c)$.

\item{2)} Unlike the case of \hw, the $N_f$ 
fourbranes connecting the D sixbranes to 
the \nsp\ fivebrane are not rigid -- they 
can fluctuate in the $(x^8, x^9)$ directions.
There are $N_f\times N_f$ complex fields 
$M_i^{\tilde i}$ $(i, \tilde i=1, \cdots, N_f)$ 
coming from 4-4 strings, parametrizing these 
fluctuations. These are Seiberg's magnetic mesons.

\item{3)} The coupling of the magnetic 
mesons $M$ to the magnetic
quarks $q$ is as expected \nati, through 
a magnetic superpotential:
\eqn\wm{W_m=M_i^{\tilde i} q^i\tilde q_{\tilde i}.}
This is clear from the geometry of open strings 
stretched between various branes. 

\item{4)} One can check that deformations of the 
electric theory agree with those of the magnetic 
one, in precisely the way described in \nati. In
particular, turning on a quark mass in the electric
description, which corresponds to moving the
D sixbranes relative to the \nsp\ fivebrane in
the $(x^4, x^5)$ directions, corresponds, after
following the path (a)-(d) above, to turning on
Higgs expectation values in the magnetic description.
Turning on an expectation value for the electric
quarks corresponds in the magnetic description
to giving the ``magnetic mesons'' $M$ an expectation
value.

\bigskip

\newsec{Duality with an adjoint superfield.}

\nref\str{A. Strominger, hep-th/9512059, 
\pl{383}{1996}{44}.}%
\nref\kk{D. Kutasov, hep-th/9512145,
\pl{383}{1996}{48}.}%
\nref\witt{E. Witten, hep-th/9512219, 
\np{463}{1996}{383}.}%
Seiberg's original work was 
generalized in \ks\ to theories 
with a single adjoint field. In \ks\ 
it was shown that the
``electric'' theory with gauge group 
$G_e=SU(N_c)$, $N_f$
flavors of fundamental matter $Q_i, 
\tilde Q^{\tilde i}$,
and a single adjoint field $X$ with 
superpotential
\eqn\wx{W_e={\rm Tr} X^{k+1},}
is equivalent at long distances to 
a magnetic theory
with gauge group $G_m=SU(kN_f-N_c)$, 
with an adjoint field
$Y$, $N_f$ flavors of magnetic quarks 
$q^i$, $\tilde q_{\tilde i}$,
and $k$ magnetic mesons $M_j$, each of 
which has $N_f^2$ components.
The full magnetic superpotential is 
(roughly -- see \ks\ for more details):
\eqn\wmag{W_m={\rm Tr} Y^{k+1}+\sum_{j=1}^k 
M_j\tilde q Y^{k-j} q.}
It is not difficult to guess the right brane 
configuration  describing this duality. 
If we replace  the single NS fivebrane of section 
2 by $k$  coincident NS fivebranes\foot{Note that 
configurations of coincident NS fivebranes appear 
in type II string  theory near $A_k$ singularities 
in the moduli space of K3 compactifications \refs{\str- 
\witt}.} all connected to a single \nsp\ fivebrane, 
and repeat the analysis of section 3, we find that 
the gauge group is transformed appropriately\foot{The
duality of \ks\ can be extended from $SU(N)$ to $U(N)$
gauge groups.}, 
$G_e=U(N_c)\to G_m=U(kN_f-N_c)$, and the right set of 
magnetic mesons $M_j$ \wmag\ appears. We will not go
through the analysis of section 3 for this case; in
deriving the result it is important to note that 
configurations where more than one fourbrane connects
the \nsp\ fivebrane to a given D sixbrane are supersymmetric. 
In contrast, in \hw\ it has been pointed out that to reproduce
the expected field theory behavior, configurations with more
than one fourbrane connecting a given NS fivebrane to a given 
D sixbrane ($s$-configurations) should not
preserve supersymmetry. Geometrically, the difference might
be due to the fact that two D fourbranes connecting  an NS
fivebrane to a D sixbrane must necessarily be on top of each other,
while different D fourbranes connecting an \nsp\ fivebrane
to a D sixbrane may be separated in the $(x^8, x^9)$ directions,
which are common to the two branes.

As discussed in detail in \ks, one way to 
understand the generalized duality is to resolve 
the superpotential \wx\ to:
\eqn\wres{W_e={\rm Tr} 
\sum_{j=2}^{k+1}a_jX^j;\;\;\; a_{k+1}=1.}
For generic values of the 
$\{a_j\}$, the bosonic potential
$V\sim |W^\prime|^2$ has $k$ 
distinct minima, in each of which
the adjoint field is massive. If 
$r_j$ of the $N_c$ eigenvalues
of $X$ sit in the $j$'th minimum 
of $V$ $(\sum r_j=N_c)$, the gauge
symmetry is spontaneously broken:
\eqn\unc{U(N_c)\to U(r_1)\times 
U(r_2)\times \cdots\times U(r_k).}
The only massless matter near 
the $j$'th minimum is $N_f$ 
fundamentals of $U(r_j)$; thus 
for generic $\{a_j\}$, the model
describes at low energies $k$ 
decoupled supersymmetric QCD theories
with gauge groups $U(r_j)$, \unc. 
One can apply Seiberg's duality
to each of the $U(r_j)$ factors, 
taking them to $U(N_f-r_j)$, as in
section 3.
Finally, in the limit $a_j\to 0$, the 
full magnetic gauge group
is restored: 
$U(N_f-r_1)\times\cdots\times U(N_f-r_k)\to
U\left(\sum_j(N_f-r_j)\right)=U(kN_f-N_c)$.

The same story can be told using the 
brane configuration described above. 
One can group the $N_c$ fourbranes
stretched between the $k$ NS fivebranes
and the \nsp\ fivebrane into groups
of $r_j$ fourbranes stretched between
the $j$'th NS fivebrane and the  
\nsp\ fivebrane. The superpotential 
\wres\ can be parametrized by the locations 
of $k$ points in a plane (the $k$ complex
solutions of $W^\prime(x)=0$). On the other 
hand, the position of each NS fivebrane is 
specified by four real numbers. Translation
of the $k$ NS fivebranes in the $(x^6, x^7)$ 
directions correspond to explicit breaking of 
$U(N_c)$ as in \unc, obtained by assigning
different gauge couplings and/or FI D-terms 
to the different $U(r_j)$. The superpotential 
is encoded in the position of the $k$ NS fivebranes 
in the $(x^8, x^9)$ plane. In fact, one can think
of the adjoint field $X$ \wx\ as the 4-4 string 
describing (infinitesimal) fluctuations
of the fourbrane in the $(8,9)$ directions.
Note that in the case where the $U(N_c)$ gauge 
symmetry is broken as in \unc, with $k$ decoupled
SQCD systems describing the different minima of
the superpotential \wres, there are parameters
in the gauge theory that can not be seen in the 
string (brane) description, corresponding to
changing the masses of the quarks around each
minimum independently (in the brane configuration,
all the masses are set to be equal). 

\bigskip

\newsec{Comments.}

\item{1)} It is clear from the brane construction
of sections 2, 3
that the electric and magnetic theories have the same
moduli space of vacua.  
This is equivalent to the statement that the two 
models have the same ring of chiral operators.
Since the models in question do not have a Coulomb 
phase, it is likely that the relation extends to 
the full infrared conformal field theory.
 
\item{2)} There was some recent work on N=1 duality
in string theory, from the point of view of F-theory 
\vvv. In the case of systems with eight supercharges, 
one can show
\ref\egk{S. Elitzur, A. Giveon and D. Kutasov, 
unpublished.}
that the analog of our approach is related by T-duality
to the study of threebranes in 
F-theory. Specifically, consider type IIB string 
theory compactified on a $K3$ surface (it is convenient 
to consider $K3\simeq (S^1)^4/Z_2$, as a starting point), 
in the presence of sevenbranes wrapping the $K3$ and 
threebranes at points on $K3$. T-duality on a single $S^1$ 
turns \kk\ this into a type IIA string compactified on a 
$T^4/Z_2$ orbifold, with sixteen NS fivebranes located 
(for generic values of the moduli of $K3$) at generic 
points  on $T^4/Z_2$. The threebranes and sevenbranes 
are transformed under T-duality into fourbranes and 
sixbranes, respectively. For the case of theories
with four supercharges, similar considerations of type IIB
on $T^6/ Z_2\times Z_2$ lead to our brane configuration.

\nref\iiss{ K. Intriligator and N. Seiberg, 
hep-th/9607207, \pl{387}{1996}{513}.}%
\nref\ooo{J. de Boer, K. Hori, H. Ooguri and Y. Oz, 
hep-th/9611063; 
J. de Boer, K. Hori, H. Ooguri, Y. Oz and
Z. Yin, hep-th/9612131.}%
\nref\pz{M. Porrati and A. Zaffaroni, hep-th/9611201.}%
\item{3)} T duality in the $x^3$ direction turns the
brane configuration presented in section 2 into a
three dimensional N=2 supersymmetric version of the 
construction of \hw. It would be interesting to use
the brane picture to study the dynamics of the
gauge theory on the threebrane, and in particular
explore the consequences of the $SL(2, Z)$ 
duality group of type IIB string theory. 
This may lead to generalizations of mirror 
symmetry \refs{\iiss, \ooo, \pz, \hw} to N=2 gauge 
theory in 2+1 dimensions.

\bigskip
\noindent{\bf Acknowledgements:} 
We thank A. Schwimmer for useful discussions.
This work is supported in part by the Israel 
Academy of Sciences and Humanities -- Centers 
of Excellence Program. The work of A. G.
is supported in part by BSF -- American-Israel Bi-National
Science Foundation. S. E. and A. G. thank the Einstein Center
at the Weizmann Institute for partial support.

\listrefs
\end